# Radiation sputtering of hydrocarbon ices at Europa-relevant temperatures

Sankhabrata Chandra, Bryana L. Henderson*, Murthy S. Gudipati
Science Division, Jet Propulsion Laboratory, California Institute of Technology, Pasadena, CA, 91109, USA
*Corresponding author: Bryana L. Henderson (bryana.l.henderson@jpl.nasa.gov)

**Highlights**

- Electron sputtering of hydrocarbons in water ice is investigated.
- The evolution of sputtered $CO_2$, CO and hydrocarbon fragments is followed.
- Sputtered species show differences in time- and temperature-dependent behavior.
- Radiolysis hampers the ability to map sputtered products to surface composition.

**Abstract**

The surfaces of some icy moons, such as Jupiter's moon Europa, are heavily bombarded by energetic particles that can alter the surface materials and affect the composition of its exosphere. Detection of $CO_2$ on Europa's surface indicate that Europa's interior may be transporting freshly exposed carbon-containing material to the surface. It is unknown whether this $CO_2$ is a product of radiation of carbon-containing precursors or whether it is present in the initial deposits. Regardless, further radiolysis by high-energy electrons or ions can sputter $CO_2$ (and organic fragments if present) into Europa's exosphere. In this study, we investigate the radiation sputtering of $CO_2$ and organic fragments from hydrocarbon water ice mixtures at different Europa-relevant surface temperatures to identify how its sputtering products evolve over time. This study shows that the sputtering of hydrocarbon water ice leads to the production of mostly $CO_2$, CO, and fragmented hydrocarbons. The onset of sputtered hydrocarbons is immediate, and quickly reaches a steady state, whereas $CO_2$ and CO are formed more gradually. It is found that higher temperatures cause more sputtering, and that there are some notable differences in the distribution of species that are sputtered at different temperatures, indicating local heterogeneity of sputtering yields depending on the surface temperature.

*Keywords: Europa, Ices, Radiation interactions (Ice), Experimental techniques, Atmospheric Composition*

# 1. Introduction

Icy moons such as Jupiter's moon Europa or Saturn's moon Enceladus are predicted to have deep, salty oceans underneath a thick layer of ice[1–5] that could reach the surface by a variety of potential mechanisms, including tectonics, impacts, cryovolcanism, or plume ejecta.[6–9] Under these conditions, material could be transported from the subsurface liquid reservoirs onto the surface. These fresh surface deposits could contain trapped complex organic molecules (COMs) in the ice or exist separately as solid COMs depending on their solubility. Indeed, the Cosmic Dust Analyzer (CDA) impact mass spectrometer onboard the Cassini spacecraft detected high-molecular weight hydrocarbons in the plumes of Enceladus, including the presence of COMs.[10] Detection of carbon dioxide ($CO_2$) on Europa's surface[11] and recent JWST observations that associate surface $CO_2$ with young chaos terrains[12] suggests an endogenic origin of $CO_2$. However, it should be noted that $CO_2$ is also known to be a common radiation byproduct and could instead result from surface irradiation of other carbon COM-like sources such as emplaced organics or carbonate-bearing minerals from the subsurface ocean.

Icy surfaces across the solar system are exposed to different levels of radiation, with some icy moons, such as Europa, experiencing higher fluxes than others.[13] Europa is constantly bombarded by high fluxes of electrons, protons and ions,[14] which can induce both physical and chemical changes that can significantly alter the composition of Europa's surface and sputtered materials can populate the Europan exosphere. Physical changes include amorphization, ejection (sputtering), and dehydration of hydrated minerals.[15–17] Chemical changes include processes such as radiolysis and recombination, and can significantly alter the surface and exospheric composition of Europa.[18] Radiolysis of COMs generally progresses toward production of more oxidized species such as $CO_2$ and organics which either can be trapped in the ice or sputtered from the surface to the exosphere. However, the detailed sputtering of COMs and how it evolves over time has not yet been studied. The sputtering process depends on several parameters including particle energy and temperature and so far, these parameters are best-studied for pure $H_2O$ ices.[19–30] Generally, the ion-induced sputtered yield for $H_2O$ ice depends on type of ion and energy of the ion, with heavier ions sputtering more efficiently than lighter ions.[20] The sputtered flux of $H_2O$ molecules has been found to be relatively constant when the ice temperature is below ~80 K, and then increases exponentially as temperatures increase toward $H_2O$'s sublimation temperature.[20,22] In contrast, yields of sputtered radiolytically-produced species like $H_2$ and $O_2$ exhibit more complex behavior, as they can be trapped inside the ice (possibly at a range of different depths depending on the stopping power of the energetic particle) and could be ejected upon subsequent warming.[25,27] These species-dependent differences indicate that different temperatures during sputtering as well as afterwards (during warming events) affect the compositional distributions of the sputtered ejecta in addition to controlling the overall ejected fluxes.

Experiments tracking sputtering products of complex ices containing other molecular additives such as hydrocarbon:water ices are less common, although the *in-situ* radiation products of electron-bombarded astrophysical ices have been relatively well-characterized using techniques such as spectroscopy and laser-desorption mass spectrometry of the irradiated ices.[31–36] One study, following the evolution of three- and four-carbon-containing species in $H_2O$ ice at 80 K upon 10 keV electron bombardment, identified many new products *in-situ* including long chain, branched aliphatic, aldehydes, ketones, esters, alcohols and $CO_2$ using IR spectroscopy of the ices and residues and temperature-programmed desorption of the photoproducts.[35] A recent study has shown the sputtering of methanol, CO, and $CO_2$ from a methanol:$CO_2$ ice matrix at 10 K by 0.57 MeV $^{58}Ni^{9+}$ ions.[37] Another study has detected carbon-containing ions being ejected from

ethane:water ices at ~60 K during bombardment with energetic multicharged heavy ions at MeV energies.[38] However, to our knowledge, there have so far been no reports of studies directly tracking electron-induced sputter products and their evolution for hydrocarbons trapped in ice.

In this work, we simulate the scenario of organics trapped in $H_2O$-ice deposits on the surface of an icy body such as Europa or Enceladus. We have chosen *n*-hexane ($C_6H_{14}$) to represent a simple COM that lacks oxygenated functional groups so that any oxidation products would be more readily identifiable. Also, the selection of n-hexane rather than any shorter hydrocarbon is based on its sublimation temperature of ~125 K.[39] Any other shorter-chain hydrocarbon may sublime on the surface of Europa as the maximum temperature of Europa is ~120 K.[51] We have tracked changes in sputtered products directly during electron irradiation to determine (1) the key sputtered species produced in irradiated hydrocarbon-containing ices that would become a part of exospheres of these icy bodies, (2) how the sputtered products change with time as the ices transition from "fresh" to "aged" by radiolysis, and (3) how the relative abundances of different sputtered products change with temperature. Our irradiation experiments are conducted at 80 K, 100 K, and 120 K, temperatures which are relevant to icy surfaces across the solar system.

## 2. Experimental Setup

The experimental setup for all the electron sputtering experiments is depicted in Figure 1 and is located at the Jet Propulsion Laboratory's Ice Spectroscopy Laboratory (ISL). The experimental setup consists of a stainless-steel vacuum chamber (purchased from Kimball Physics, Inc., USA), an electron gun (Model: ELG-2/EGPS-1022, Kimball Physics, Inc., USA), and a mass spectrometer (RGA200, Stanford Research Systems, Inc., USA). The vacuum chamber pressure is monitored by a hot cathode ionization gauge (Model: KJLC354401YF, Kurt J. Lesker Company, USA) which can operate from $6.6 \times 10^{-2}$ mbar to $1.3 \times 10^{-9}$ mbar, and the vacuum is maintained by an 80 liter/second turbomolecular pump (Model: TwisTorr 84FS, Agilent Technologies, USA) that is backed by a dry scroll pump. With this pumping system the vacuum is maintained to better than $10^{-8}$ mbar. The temperature of the sample holder is maintained by a closed-cycle helium cryostat (Advanced Research Systems, Inc., USA). A copper rod with surface area of ~100 $mm^2$ is used on the top of the cold head as a sample substrate. The temperature of the top of the sample holder is controlled using two programmable cryogenic temperature sensors (DT-670 silicon diode, LakeShore Cryotronics, Inc., USA). One is placed at the bottom of the sample holder and the other is located 10 mm below the sample holder's top. For preparing the samples, degassed, ultrapure water (JT Baker Chemicals, Inc., USA) and hexane (HPLC grade, Fisher Scientific Co LLC) are vapor deposited through two separate leak valves. Before vapor deposition three freeze-pump-thaw cycles are performed at a pressure of ~$10^{-8}$ mbar to remove any unwanted dissolved gases in the liquids. A few microns thick $H_2O$:hexane ice is formed by deposition of $H_2O$ and hexane simultaneously for 30 minutes at a pressure of $5 \times 10^{-7}$ mbar based on earlier measurement by laser interferometric method.[36,40] To sputter $H_2O$:hexane ice an electron energy of 2 keV is used with an electron current of 5 microamperes. The focus and grid voltages are maintained at 0.2 V and 0.80 V respectively. The angle of the electron gun is 45 degrees with respect to the normal of the ice surface formed on the sample holder. After considering the geometrical arrangement of the electron gun, electron energy, electron current, voltage of focus and grid, the total energy flux on the ice surface is about $1.1 \times 10^{13}$ keV $cm^{-2}$ $s^{-1}$. For comparison, Europa's global electron energy flux has been given as $6.2 \times 10^{10}$ keV $cm^{-2}$ $s^{-1}$ based on the EPD measurements from the Galileo Orbiter.[13,18,41] Using this value, two hours of radiation in our experiment is roughly equivalent to 356 hours on Europa. However, this is likely an overestimate of the number of equivalent hours on Europa, as the Europa energy flux number above includes only electrons in the 20 keV to 40 MeV range. The inclusion of electrons outside this range, or of other energetic particles such as $H^+$, $O^{n+}$, or $S^{n+}$ ions would lead to a higher overall Europa energy flux.

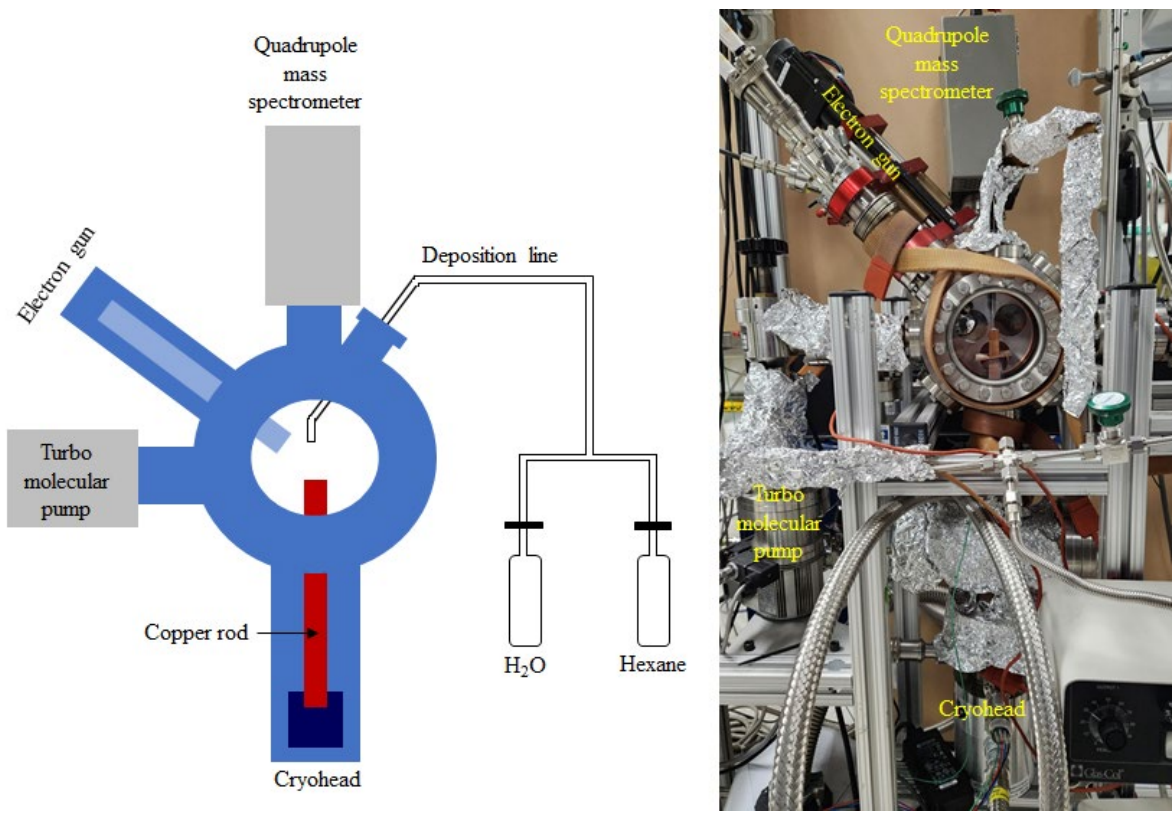

Figure 1: Experimental setup for the electron sputtering experiments of ice surfaces, installed at the Jet Propulsion Laboratory's Ice Spectroscopy Laboratory (ISL). Left: the vacuum chamber with electron gun and quadrupole mass spectrometer. $H_2O$ and hexane are vapor deposited on the top of the cold copper rod. Right: a photograph of the system.

## 3. Results and discussion

The purity of the hexane sample was confirmed by introducing it as a vapor at room temperature into the vacuum chamber with a pressure of $1\times10^{-6}$ mbar and taking a quadrupole mass spectrometer (QMS) measurement using a filament ionization energy of 70 eV (Figure 2 inset). The fragmentation pattern and intensities match well with a hexane mass spectrum from the NIST database[42] (shown in green), which was also collected using electron impact ionization. All species are detected as cations in the QMS in all experiments. Most of the hexane fragmented to three- and four-carbon components, and the molecular hexane ion intensity at m/z 86 is low in comparison.

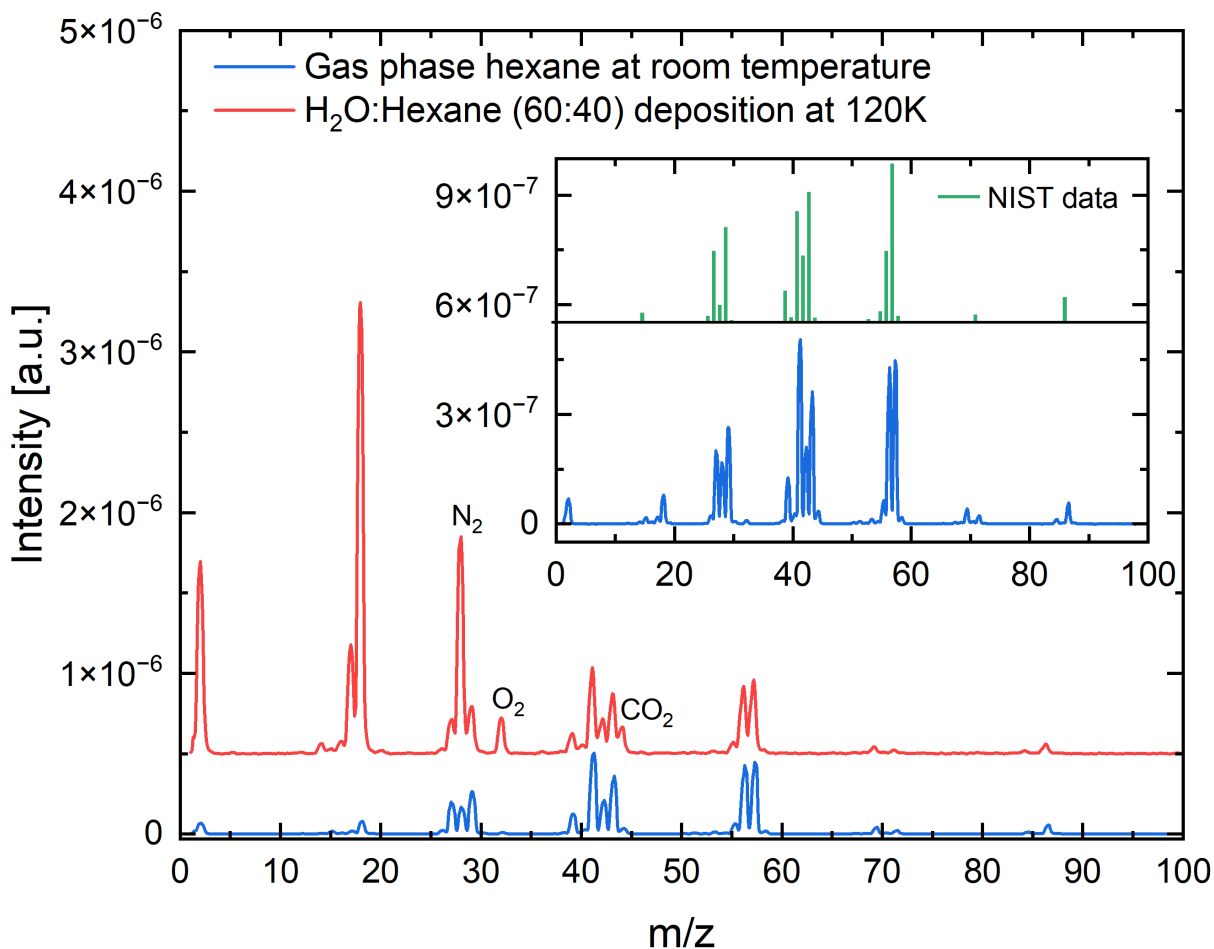

Figure 2: Scaled plot with respect to hexane is shown for the gas phase hexane and during deposition. During deposition, the dissolved gases in $H_2O$ are clearly visible as $N_2$, $O_2$ and $CO_2$ in the mass spectrum (red). Inset shows the fragmentation pattern of gas phase hexane and NIST data for hexane at room temperature.

To form the ice analogs, $H_2O$ and hexane are introduced through two separate dosing valves and combined into a single line before finally being deposited on the copper surface at 100 K. To prepare ~60:40 (mol%) $H_2O$:hexane ice, partial pressures of $H_2O$ and hexane are maintained at ~$3\times10^{-7}$ mbar and ~$2\times10^{-7}$ mbar, respectively on the ion gauge. This is achieved by introducing $H_2O$ first in the vacuum chamber followed by hexane to maintain a total pressure of $5\times10^{-7}$ mbar. The vapor pressures of $H_2O$ and hexane were meticulously regulated, permitting a maximum of 5% variation in the intended composition. The $H_2O$:hexane ratio is further confirmed by the measurement from QMS during deposition. Hexane contribution is determined by integrating all the fragments expected to have more than 10% contribution according to NIST EI mass spectrometer data.[42] We used a relatively high proportion of hexane (40% by mole) to ensure that sputtered hydrocarbons and products are detectable by the mass spectrometer. Figure 2 shows the mass spectra for the gas phase hexane and the $H_2O$ and hexane mixture. Small amounts of residual $N_2$, $O_2$ and $CO_2$ (which are shown in Figure 2) are also present, and some of these species are deposited or trapped in the ice during deposition despite the 80-120 K temperatures. To remove the sputtered contribution from the inadvertently trapped species, a separate sputtering experiment was conducted with pure $H_2O$ ice and the mass spectra were subtracted from all $H_2O$:hexane sputtering experiments.

During each experiment, the system is kept under vacuum for 120 minutes to achieve a stable background pressure in the range of $\sim 1\times 10^{-9}$ mbar before electron sputtering started. The ice is sputtered with a flux of $1.1\times 10^{13}$ keV electrons $cm^{-2} s^{-1}$. The mass spectra of gases in the chamber are averaged for 120 minutes of before sputtering and during sputtering and are plotted in Figure 3 in the top and middle panel, respectively, with the ice held at 120 K. An increase of different sputtered species can be seen in the middle panel of Figure 3. The increase of the sputtered species is more obvious in the bottom panel which shows the subtraction of top spectrum (before sputtering) from the middle spectrum (during sputtering). A discontinuity at m/z=32 is observed due to the depletion of gas-phase background $O_2$ due to radiolysis in the gas-phase during sputtering, as a result becoming a negative value in the logarithmic plot. In pure $H_2O$ ice, $O_2$ is a common sputtered byproduct.[35] However, $O_2$ was not clearly observed in our experiments. In our 40% hexane/60% water ice (as measured by partial pressure during deposition in the mass spectrometer), the C/O ratio in the ice is 4:1. With this ratio, oxygen radicals may be statistically more likely to react with carbon fragments rather than another atomic oxygen, producing $CO_2$ and CO rather than oxygenated species such as $O_2$ or $O_3$. Peaks are present at m/z=28, 32 and 44 in all three panels. Previously a number of studies reported that the sputtering of $H_2O$ ice produces sputtered species such as $H_2$, O, OH, $H_2O$, $O_2$, $H_2O_2$ and $O_3$.[43–48] The combination of $H_2O$ and hexane mixed ice can additionally produce a number of sputtered species by reaction between different radical species coming from $H_2O$ and hexane in the ice. This can lead to the production of oxidized species such as alcohols, aldehydes, acids, CO and $CO_2$. However, to identify every species in the mass spectrum is challenging since most of the species overlap with hydrocarbon fragments and oxygenated hydrocarbon species. Therefore, in this work, we only track the most stable oxidation products ($CO_2$, CO) along with a few hydrocarbon mass peaks that are clearly identifiable during the course of the sputtering process. To confirm the identification of the molecules/fragments with their mass peaks, a separate sputtering experiment with isotopically-labeled $H_2^{18}O$ with hexane ice was performed at 120 K keeping all the other parameters the same as with previous experiments. The sputtered species up to m/z=100 are compared in Figure 4 for the $H_2^{18}O$:hexane ice and $H_2O$:hexane ice experiments. In the isotope-enriched experiment, 97% $H_2^{18}O$ (Sigma-Aldrich, USA) is used as shown in Figure S1 bottom panel. The mass spectrum of gas phase normal ($^{16}O$-isotope) $H_2O$ is also shown in the top panel. Using the same deposition settings, the isotope-labeled water (bottom panel) deposits approximately 28% of normal $H_2^{16}O$ (m/z=18) with respect to $H_2^{18}O$. Therefore, some contribution from normal $H_2O$ (possibly from gas phase water present as background during deposition) is seen in the isotope-enriched experiments. In the following sections, identification of m/z of 15, 29, 43, 28 and 44 are presented through comparison of $H_2O$:hexane ice and $H_2^{18}O$:hexane ice experimental results.

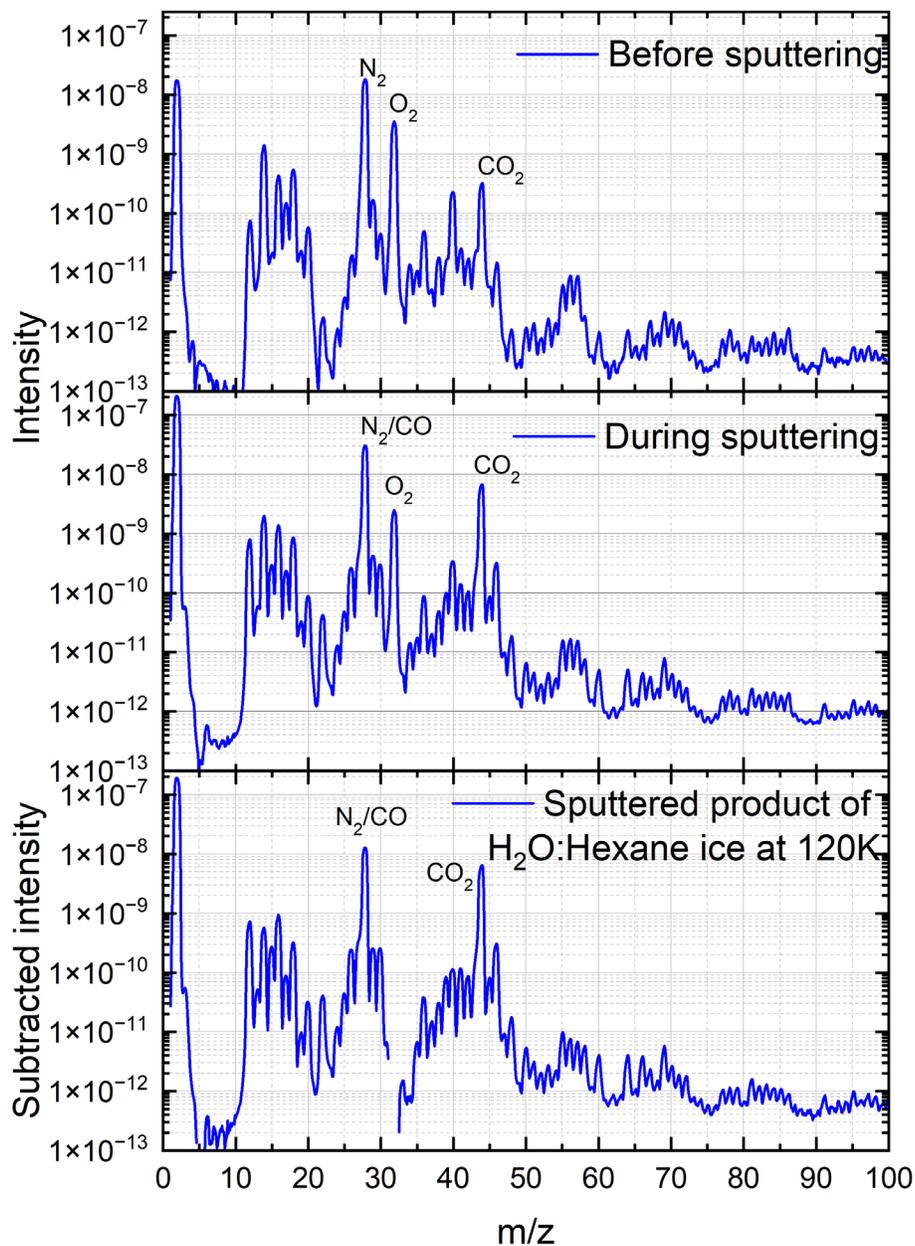

Figure 3: Mass spectra of $H_2O$:hexane ice before sputtering (top) and during sputtering (middle), each integrated for 120 minutes. The bottom panel represents the subtraction of before sputtering from during sputtering mass spectra of $H_2O$:hexane ice at 120 K. A discontinuity is observed at m/z 32 due to the depletion of background $O_2$ when the electron source is turned on.
7

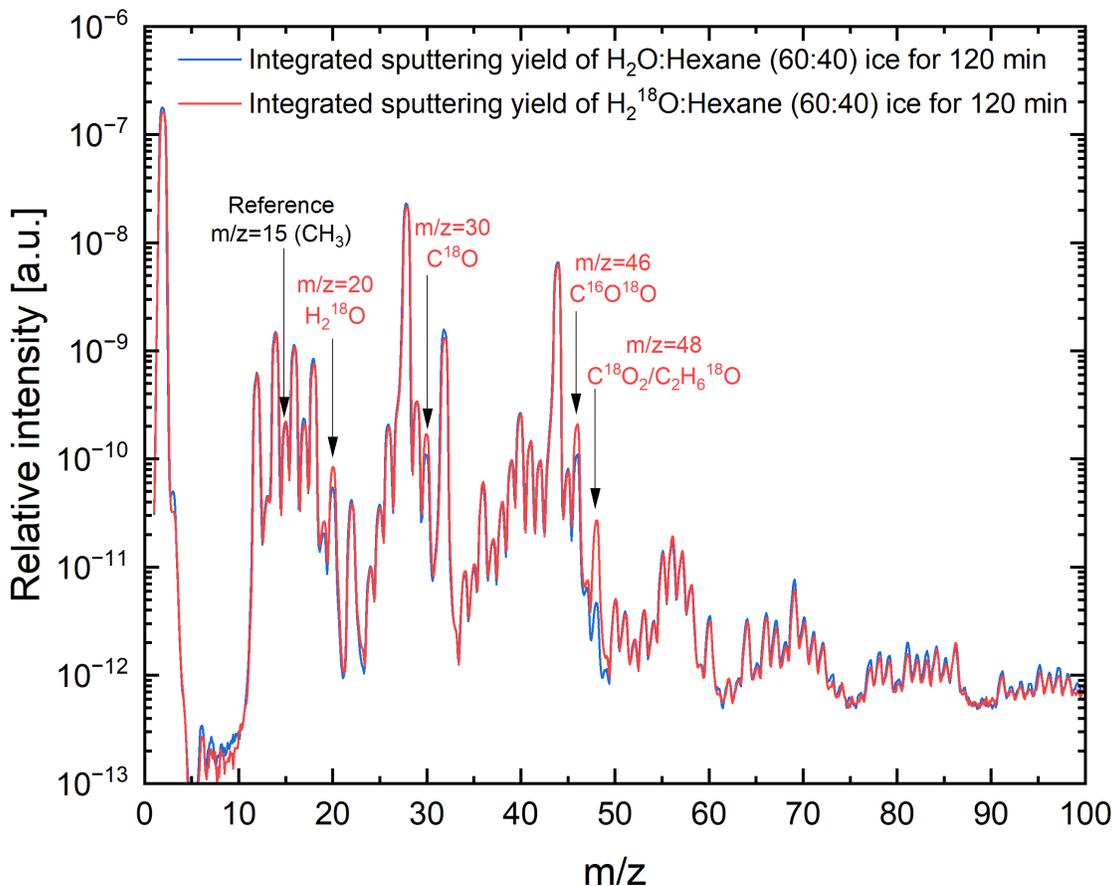

Figure 4: Comparison of mass spectra of the $H_2O$:hexane and $H_2^{18}O$:hexane sputtered ice, integrated over 120 minutes of electron bombardment at 120 K temperature. The $H_2O^{18}$:hexane sputtered ice spectrum is scaled with respect to the m/z=15 peak and is not background subtracted. The spectrum also shows the shift of m/z= 44 to 46 and 48 and m/z=28 to 30. This shows $^{18}O$ is reacting with hydrocarbon fragments to produce $C^{18}O_2$ and $C^{16}O^{18}O$ and $C^{18}O$.

## 3.1 Assignment of key species

**m/z=15 ($CH_3$)**

A plot of the region from m/z=10 to 20 is shown in Figure 5a, comparing the background (before sputtering) and the sputtered (during sputtering) signal from $H_2O$:hexane ice, both integrated for 120 minutes at 120 K temperature. Before sputtering begins, a small amount of m/z=12 and 14 are observed and assigned to atomic carbon and atomic nitrogen (from residual $N_2$ contamination during deposition). m/z=17 and 18 are assigned to OH and $H_2O$ since the ice contains 60% of $H_2O$. During sputtering there is an increase of all peaks between m/z 12 and 18. Although a small portion of m/z 15 could correspond to NH, we assign the majority of this peak to the hydrocarbon fragment $CH_3$. This signal also exhibits a sputtering profile with time that is similar to other hydrocarbon fragment peaks, and this will be discussed further in the "sputtering evolution of hydrocarbon fragments" section.

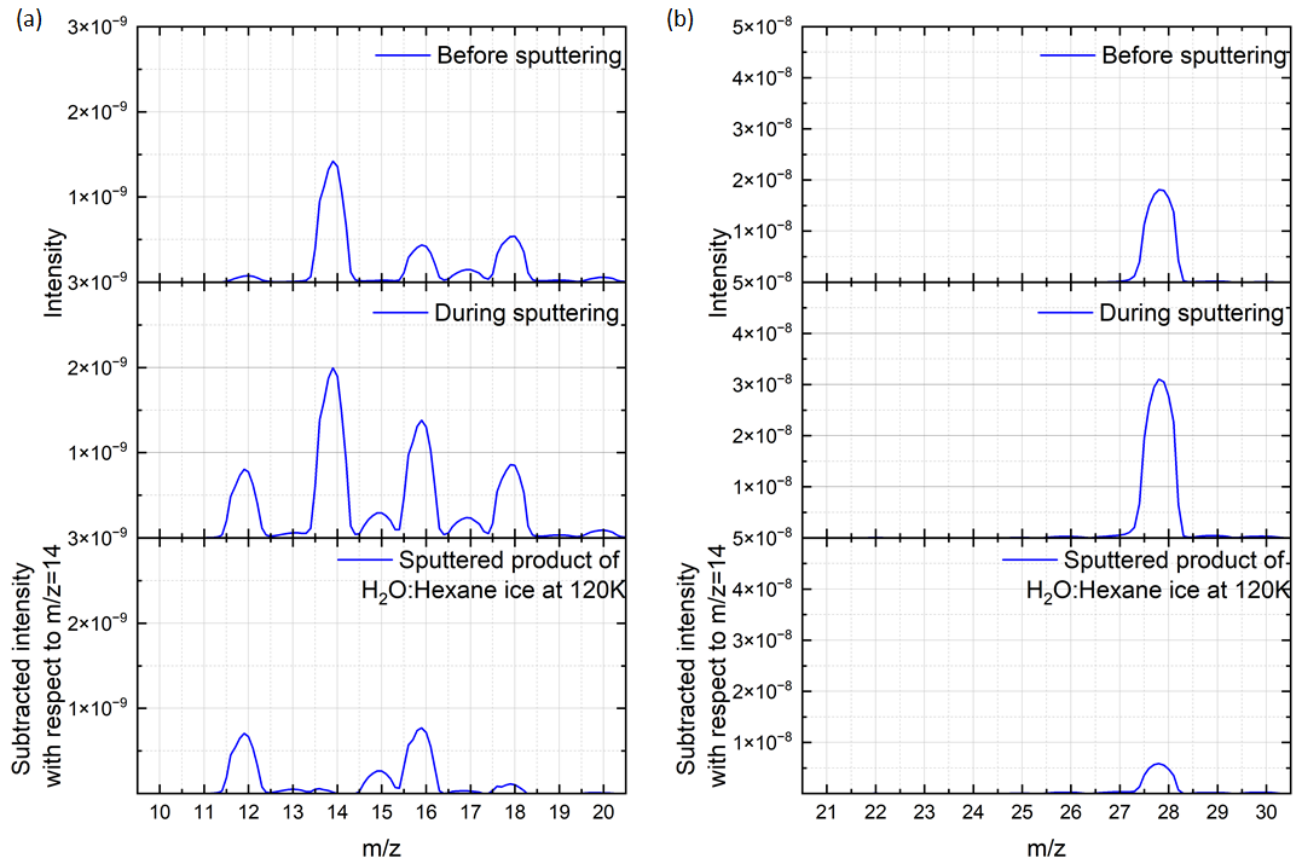

Figure 5: Integrated mass spectra of $H_2O$:hexane ice are shown in the top and the middle panels before sputtering and during sputtering, respectively, for 120 minutes in the range of (a) m/z=10 to m/z=20 and (b) m/z=21 to m/z=30. The bottom panel represents the subtraction of before sputtering from during sputtering mass spectra of $H_2O$:hexane ice at 120 K after normalizing with respect to m/z=14 to remove the contribution of N and $N_2$. (a) and (b) represent the same plots using a different 'y' scale.

**m/z=29 ($C_2H_5$)**

Mass to charge ratio of 29 can be assigned to either a hydrocarbon fragment $C_2H_5$ or the oxygenated species HCO. To identify m/z=29, we compared $H_2O$:hexane ice to $H_2^{18}O$:hexane ice results. Figure 6 top and middle panels represent the first and last 60 minutes (each integrated) of the sputtered species from m/z=21 to m/z=50. A full-scale plot of these panels is shown in Figure S2. No clear shift of m/z of 29 to 31 was observed while comparing the integrated signal of $H_2O$:hexane ice and $H_2^{18}O$:hexane ice, while an increase in m/z 30 due to $C^{18}O$ is clearly seen in the $H_2^{18}O$ ice experiments, which is expected. These experiments suggest that the majority of the peak at m/z 29 is due to a non-oxygen containing species such as $C_2H_5$.

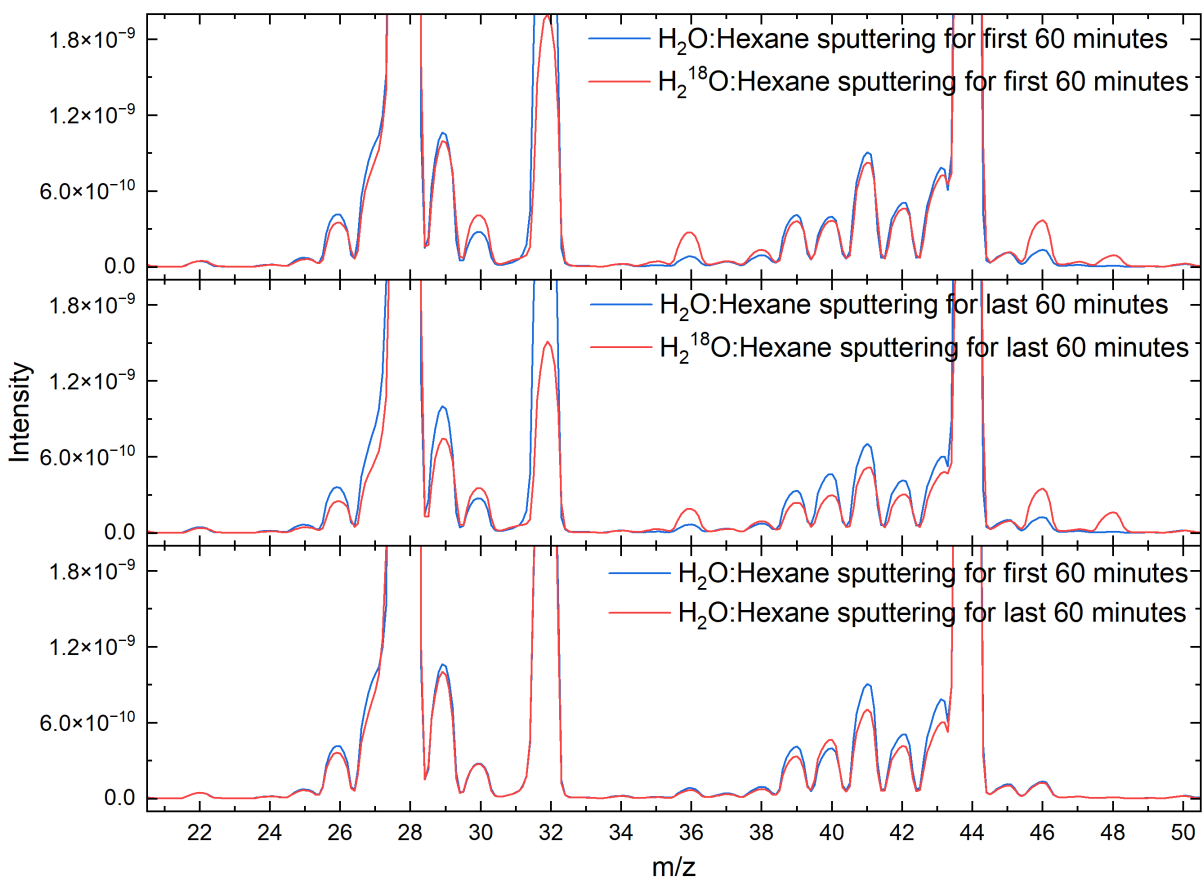

Figure 6: A comparison of the integrated, non-background-subtracted sputtered intensity for the mass to charge ratio of 21 to 50 is plotted for $H_2O$:hexane and $H_2^{18}O$:hexane ice for the first and last 60 minutes at 120 K in the top and middle panel, respectively. An increase in m/z 30 due to $C^{18}O$ is observed in the $H_2^{18}O$ ice experiments (relative to the non-labeled experiments, but no clear transition from 29→31 or 43→45 m/z is seen, indicating that m/z 29 and m/z 43 are predominantly representing non-oxygen-containing species. The non-background-subtracted first and last 60 minutes of the $H_2O$:hexane ice at 120 K temperature is also compared in the bottom panel.

**m/z=43 ($C_3H_7$)**

Like m/z 29, identification of 43 could be attributed to either a hydrocarbon $C_3H_7$ or an oxygen-containing species $C_2H_3O$. To determine how much contribution could be coming from each source, we again compared the mass spectra resulting from sputtering of $H_2O$:hexane ice with that of $H_2^{18}O$:hexane ice as shown in Figure 6 top and middle panel for the first and last 60 minutes. In the first 60 minutes, the sputtering intensity of m/z=43 is comparable for both sets of data, suggesting that the initial m/z 43 peak is primarily due to a hydrocarbon. The last 60 minutes shows a slight decrease of the intensity of m/z=43 for the $H_2^{18}O$:hexane ice, but no corresponding increase in m/z=45 peak ($C_2H_3^{18}O$) was seen, so we assign the majority of the m/z=43 peak to the hydrocarbon fragment $C_3H_7$.

**m/z=28 ($N_2$ or CO or $C_2H_4$)**

Assignment of the peak at m/z=28 to CO or $C_2H_4$ was challenging because of the presence of residual $N_2$ in the chamber. Figure 5a and Figure 5b depict the m/z=10 to 20 and m/z=20 to 30 regions for the before sputtering and during sputtering in the top and middle panel, respectively. A subtraction plot, normalized to m/z=14, is also shown for both of the Figures. The amount of CO (produced by radiolysis and sputtering) and the amount of CO from $CO_2$ are estimated to be ~4x and ~100x smaller than the contribution of $N_2$, respectively, based on Figure 5b top and bottom panels and an estimation using the NIST fragmentation ratios of $CO_2$,[42] so we consider $N_2$ to be the primary contributor to the m/z 28 peak in the residual gas. Based on the ratio of 28 to 14 m/z, it appears that the molecular nitrogen contribution is 12.5 times higher than the atomic nitrogen contribution as shown in Figure S3 (top) in Supporting Information. This factor was used to remove the molecular nitrogen contribution from CO in the sputtered species which is shown in Figure S3 (middle and bottom). The final CO contribution is obtained after subtraction of both $N_2$ and $H_2O$ background contributions. This assignment is corroborated by the isotopic shift of two mass units in Figure 4 (from m/z of 28 to 30). Though m/z at 27 and 26 could be due to fragmentation of $C_2H_4$ (m/z at 28), the presence of these peaks already in background (before sputtering) and the fact that there are no significant changes in their peak intensities suggest that $C_2H_4$ may not be a significant contributor to the m/z=28.

**m/z=44 ($CO_2$)**

m/z=44 is identified by comparing the $H_2O$:hexane ice and $H_2^{18}O$:hexane ice sputtered data from Figure 4. This shows a shift of m/z=44 to m/z=46 and m/z=48, due to incorporation of $^{18}O$. As stated previously, a small amount of $^{16}O$ is also present in the deposited ice and this leads to the formation of both $C^{16}O^{18}O$ and $C^{18}O_2$. m/z=46 from $H_2^{18}O$:hexane ice increased with respect to m/z=46 from $H_2O$:hexane ice from $1 \times 10^{-10}$ mbar to $2 \times 10^{-10}$ mbar and m/z=48 from $H_2^{18}O$:hexane ice increased with respect to m/z=46 from $H_2O$:hexane ice from $3 \times 10^{-12}$ mbar to $2 \times 10^{-11}$ mbar.

Other than the assigned masses above, there are possibilities of sputtering of other oxygenated species such as aldehydes, alcohols, ketones, and esters. These species have been observed previously by Hand and Carlson when sputtering butane:water ice at 70 K.[35] In this work, m/z=31 and m/z=45 could potentially represent $CH_3O$ and $C_2H_5O$, although clear shifts of m/z=31 to 33 and 45 to 47 are not observed. The intensities of these masses are hard to evaluate due to nearby high intensity masses m/z=32 and m/z=44.

### 3.2 Sputtering evolution of hydrocarbon fragments

Figure 7 (top) represents the sputtered evolution of the fragmented hydrocarbon species. Small hydrocarbon fragments such as m/z of 15 ($CH_3$), 29 ($C_2H_5$) and 43 ($C_3H_7$) evolve in a similar way. These fragmented hydrocarbon masses show a sharp rise followed by a slower rise in their mass intensities that reach a maximum at 30 minutes and then the yields drop. After 60 minutes, the signal appears to reach an equilibrium. The observed trend could be due to increasing signal as the ice surface populates with fragmented hydrocarbon chains, followed by a decrease as atomic oxygen from the $H_2O$ ice starts to react with the hydrocarbon fragments.

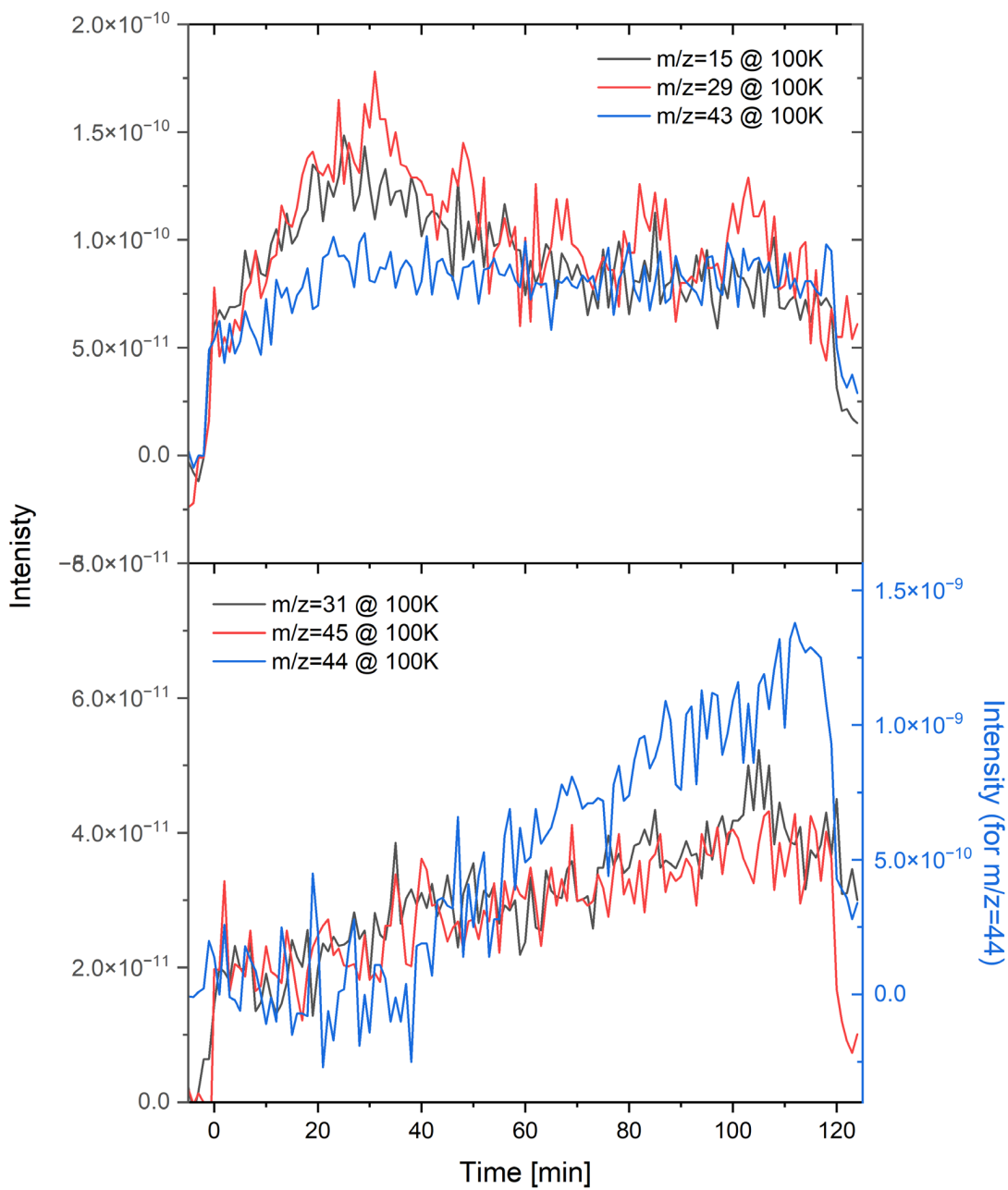

Figure 7: Top panel represents the sputtered evolution of likely hydrocarbon-based species for m/z of 15, 29 and 43 in the $H_2O$:hexane ice at 100 K. Bottom panel tracks the likely oxygenated species for m/z of 31, 45 and 44 in the $H_2O$:hexane ice at 100 K. The evolution profile shows fast production of hydrocarbon fragments and slower production of oxygenated species.

### 3.3 Sputtering evolution of oxygenated components

Although m/z=31 and m/z=45 components were difficult to definitively assign from the isotopic data, similarities in their evolution profiles suggest that a portion of these peaks may be attributed to $CH_3O$ and $C_2H_5O$. At 100 K, these peaks follow a similar pattern to $CO_2$, as shown in the bottom panel of Figure 7 (right axis), where they rise slowly and steadily compared to hydrocarbon fragments and they do not exhibit clear maxima in the 120 minutes sputtering time. Although other peaks were present that could correspond to known radiation products[36,49,50] (e.g., formaldehyde, acetaldehyde, formic acid), several overlapped with other possible fragments and we could not conclusively identify any acid or aldehyde species from shifts in the isotope experiment spectra.

### 3.4 Time- and temperature- dependent sputtering

Depending on location, season, and time of day, icy bodies such as Europa experience temperature changes that affect the sputtering profiles. To study the time- and temperature-dependence of the sputtering process, we investigated how each sputtered species evolves for three temperatures: 80 K, 100 K and 120 K, spanning a range seen at the equator of Europa over a typical day,[51] and over the course of two hours of accumulated radiation dose. Figure 8a shows the evolution for the hydrocarbon components at m/z 15, 29, and 43 (assigned to $CH_3$, $C_2H_5$ and $C_3H_7$) at the three temperatures over the course of two hours. As expected, it is found that higher temperatures give higher intensities of sputtered hydrocarbon products. In 80 K spectra, there is no notable change with time (other than a small jump when the sputtering is first initiated). All three hydrocarbon fragmented components ($CH_3$, $C_2H_5$ and $C_3H_7$) exhibit a similar pattern for 100 K and 120 K temperatures, with maxima around 30 minutes and the hydrocarbon species tailing off at longer irradiation times. In contrast, the peaks assigned to oxygen-containing species ($CH_3O$, $C_2H_5O$, and $CO_2$) shown in Figure 8b appear very shortly after irradiation starts at 100 K and 120 K and do not exhibit any distinct maxima. We interpret this as evidence of rapid transformation from the initial hydrocarbon to more oxidized species such as $CO_2$ that are very volatile at 120 K and can rapidly escape to the vapor phase, leaving behind an ice with an increasing C/O ratio. Eventually, the sputtered concentrations of hydrocarbon fragments begin to build until they themselves are depleted or converted into more refractory species.

To quantify how the distribution of sputtered species changes with respect to temperature, integrated intensities are plotted in Figure 9 for each species for three cases: the total sputtering time of 120 minutes, first 60 minutes, and the last 60 minutes (the numerical values for these can be found in Table S1). As it has been reported previously by other groups,[20,22] the sputtered $H_2O$ shows a roughly exponential increase with temperature, relative to the sputtered species observed at 80 K, in the first 60 minutes. However, the last 60 minutes of sputtering data suggest that this temperature dependence for water is less prominent than in fresh samples. The temperature dependence of the different masses shown in Figure 9 likely is a confluence of several factors, as the higher temperatures may enhance mobility in the ice and increase reaction rates but also increase efficiency of sputtering for lower-volatility products. The general trends seem to be as follows: (a) when a species is present in the initial ice or formed from simple bond-breaking, then its temperature dependence is related to its desorption temperature, and (b) if the species is formed from the combination of two or more radicals, then it appears that it requires higher temperatures or higher ice mobility to form, and then is either sputtered directly or according to its volatility.

For example, m/z=31 for $CH_3O$ is formed by combination of $CH_3$ and O radicals and this becomes more feasible when radicals have higher ice mobility. For CO and $CO_2$, more bonds need to be broken to make them in the first place, but they are also very volatile at higher temperatures and are lost easily once formed. Even in simple ice mixtures like ours, the changes in distribution of sputtered species over time and with temperature makes it difficult to use the sputtered species to infer the initial surface composition. On the surface of a radiation-drenched moon such as Europa, oxidation induced by radiolysis can obscure the identity of any biomolecules that are originally present.

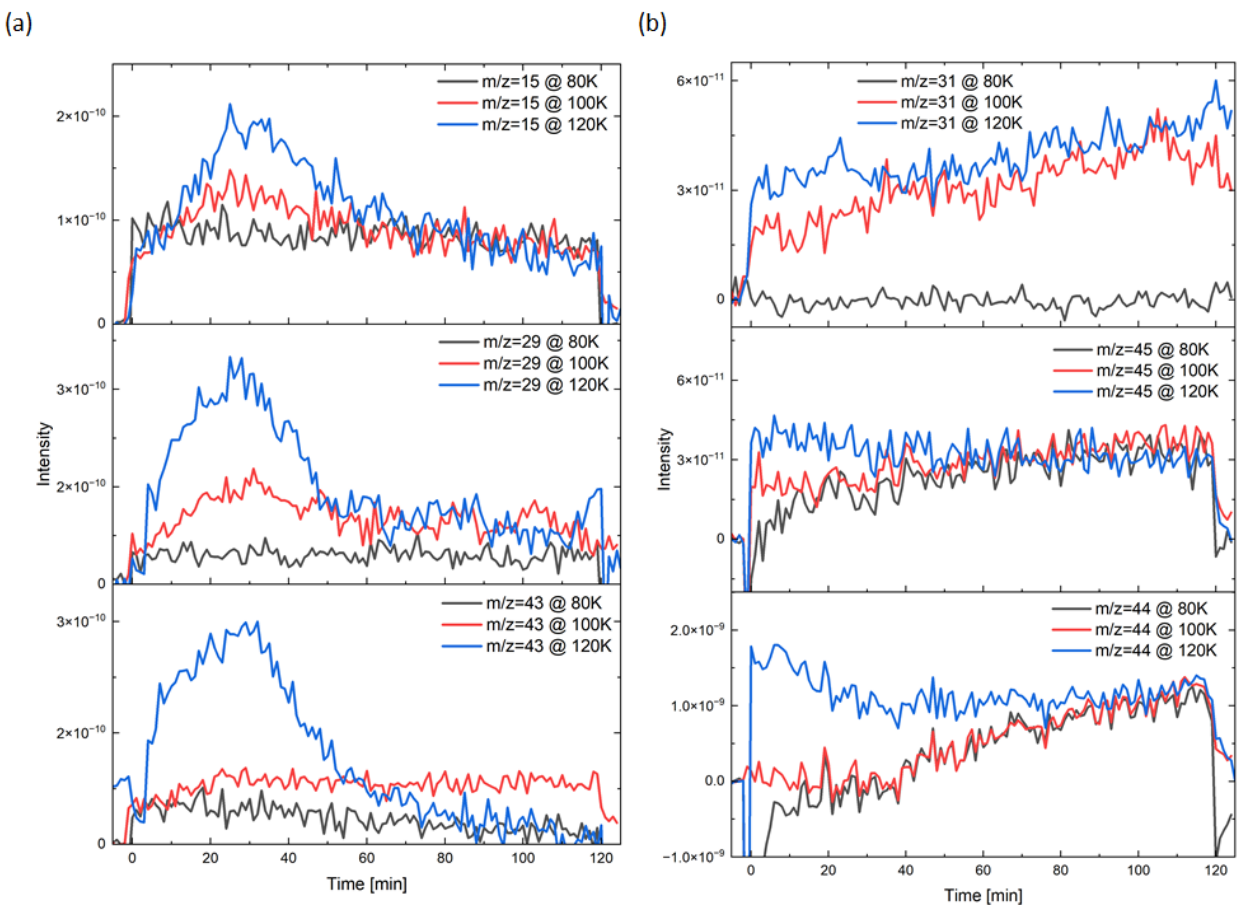

Figure 8: Sputtered evolution of (a) fragmented hydrocarbons for m/z of 15 ($CH_3$), 29 ($C_2H_5$) and 43 ($C_3H_7$) (b) possible oxygenated species for m/z of 31 ($CH_3O$), 45 ($C_2H_5O$) and 44 ($CO_2$) for three different Europa temperatures. Higher intensities of sputtered masses are obtained at higher temperatures, with temperature making more of a difference for certain species (such as m/z 31).

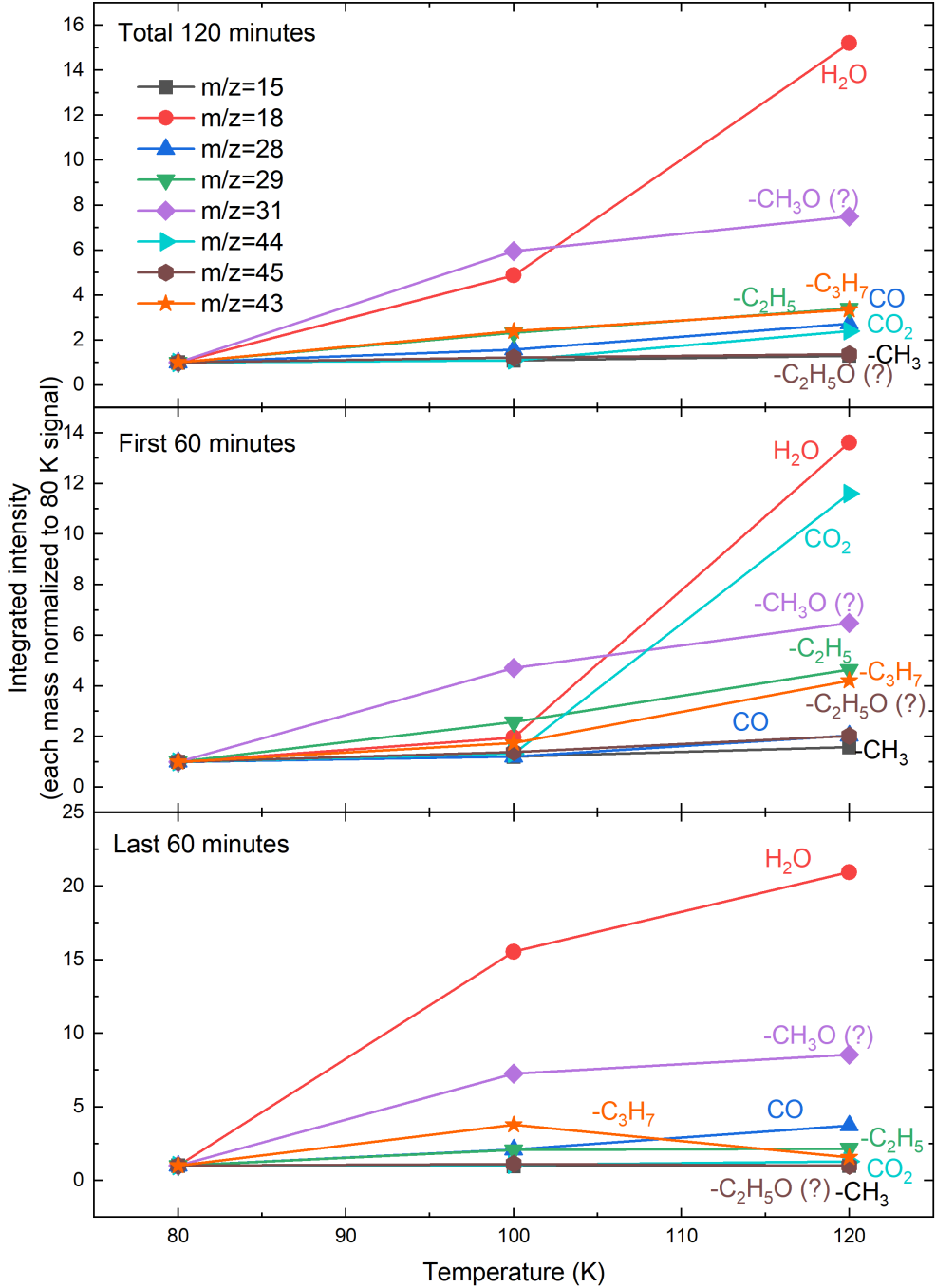

Figure 9: Sputtered species are shown at three different Europa temperatures, 80 K, 100 K and 120 K, integrated for the whole 120 minutes (top), first 60 minutes (middle) and the last 60 minutes (bottom). Each species is normalized with respect to the 80 K integrated intensity.

## 4. Conclusion

This work investigates how electron sputtering of hydrocarbons (here *n*-hexane) generates oxygenated products that can then be sputtered into the exospheres of icy bodies such as Europa. Here, a mixture of water ice with hexane (60:40) mixture is used to understand the evolution of sputtered species at different temperatures. We find that production of $CO_2$ is significant and that hydrocarbon fragments are also directly sputtered into the vacuum chamber (equivalent to Europa's exosphere). We also find some evidence for the formation of oxygenated hydrocarbons such as $CH_3O$ and $C_2H_5O$. We find the hydrocarbon fragments are immediately produced upon electron-irradiation and reach a maximum then decrease in abundance, reaching a steady-state, whereas oxygenated species gradually increase with sputtering exposure. This is in line with the logical expectation of how many chemical bonds needs to be broken or formed in the ice to generate the sputtered volatiles. Smaller hydrocarbon fragments need the least number of reactions (bond-breaking) and hence are formed rapidly upon receiving energy from electron radiation. Following the same logic, formation of oxygenated volatiles and $CO_2$ takes several reaction steps, so these sputtered volatiles are continuously generated under electron-irradiation for as long as a source of carbon is present. The timing of the onsets seen in Figures 7-8 could therefore represent differences in radiation chemistry on the surface of ice, where rapid onsets represent products that are able to be formed quickly and are lost from the surface. The sputtering also exhibits temperature-dependent behavior, where a higher temperature causes an increase in sputtered species, particularly for $H_2O$ and $CH_3O$ relative to the 80 K temperature experiment. Since the equatorial regions of Europa's surface, particularly during the daytime, have higher temperatures than the polar and night time areas, it is likely that the maximum amount of complex sputtered species would be found in Europa's exosphere during daytime near the equatorial regions. Low-volatility products generated during radiation bombardment could build up in the ice and be released as the surface rotates and is warmed by the sun. Hence, we can expect high diurnal variation in sputtered volatiles (including $CO_2$) in the equatorial regions, whereas in the polar regions these volatiles could be trapped for longer periods before other processes such as micrometeorite gardening release them into the exosphere.

## 5. Acknowledgements:

This research was carried out at the Jet Propulsion Laboratory, California Institute of Technology, under a contract with the National Aeronautics and Space Administration. S.C., B.H. and M.S.G. acknowledge support from NASA's Habitable Worlds and Solar System Workings Programs.

## 6. Copyright statement